\documentclass[12pt]{iopart}
\usepackage{graphicx}

\begin{document}

\title{Does the empirical meson spectrum support the Hagedorn conjecture?}

\author{Thomas D Cohen and Vojt\v{e}ch Krej\v{c}i\v{r}\'{i}k}

\address{Maryland Center for Fundamental Physics
Department of Physics,\\ University of Maryland, College Park, MD 20742-4111}

\eads{\mailto{cohen@physics.umd.edu}, \mailto{vkrejcir@umd.edu}}

\begin{abstract}
It has long been conjectured that strong interactions give rise to a Hagedorn spectrum and theoretical arguments have been presented in support of Hagedorn spectrum in large $N_c$ QCD.  This paper  discusses the extent to which the meson spectrum should be viewed as evidence for  a Hagedorn spectrum and argues that data do not provide a strong evidence for the Hagedorn  conjecture. The conclusion is based on three reasons.  It is shown that   ``realistic'' quark models have a spectrum in which the number of mesons up to 2.3 GeV grows with mass in a very similar way to the spectrum of physical mesons   up to 2.3 GeV.    However, these models can be shown not to have Hagedorn spectra.    It is also shown that the available data are insufficient to determine the Hagedorn temperature. The data can be described with comparable accuracy by various  functional forms of the prefactor that yield radically different Hagedorn temperatures.  An  analysis of the behavior of the spectrum for the various parity-spin-charge conjugation-isospin channels also appears to be inconsistent with what one  expects if the data were in the regime dominated by exponential behavior.
\end{abstract}




\submitto{\jpg}
\maketitle

\section{Introduction}

The Hagedorn conjecture that the density of hadrons in the spectrum grows exponentially with the mass of the hadron is now more than 45 years old \cite{Hagedorn1, Hagedorn2}.   It has played an influential role in the development of particle physics with implications ranging from QCD thermodynamics \cite{HuangWeinberg, CabiboParisi, Cohen2006} to string theory \cite{Strings}.   At a practical level, it is invoked in present day analysis of ultrarelativistic  heavy ion collisions in the context of resonance gas models \cite{BraunMunzinger1, BraunMunzinger2, BraunMunzinger3}.   Given the significance of the Hagedorn spectrum, it is important to ascertain whether  the conjecture is, in fact, correct.

Over the years, several theoretical arguments have been developed in support of Hagedorn conjecture beyond Hagedorn's phenomenological motivations.  One argument is that one expects highly excited mesons in QCD to form flux tubes \cite{IsgurPaton} and the dynamics of  long flux tubes are naturally stringlike.  Since string theories are known to have Hagedorn spectra \cite{Strings} it is natural to assume that Hagedorn spectra exist for mesons, too.  There is a number of other theoretical reasons to believe that QCD has a Hagedorn spectrum.  For example, an explicit calculation of the spectrum in QCD with adjoint fermions in 1+1 dimensions indicates a Hagedorn spectrum \cite{KoganZhitnitsky}  as does a semi-classical analysis of ${\cal N}=1  $ super Yang-Mills using gauge gravity duality \cite{PandoZayas}.   These arguments are quite powerful.  However, they suffer from important limitations.  The first argument is based on an important dynamical assumption, namely that the dynamics really is dominated by stringlike degrees of freedom.  The latter two arguments are based on theories which differ in fundamental ways from QCD.  They both have fermion field content  different from QCD, and in one case, a different space-time dimension.  Recently, an indirect argument for a Hagedorn spectrum was developed for  QCD in 3+1 dimensions with the correct fermion field content  \cite{Cohen2010, CohenKrejcirik} using properties of a matrix of point-to-point correlation functions of composite operators.

 While the argument of Refs.~\cite{Cohen2010, CohenKrejcirik}  avoids some of the problems of the other approaches, it shares with them a common limitation: it is  based---either implicitly or explicitly---on large $N_c$ QCD rather than QCD with $N_c=3$.  Consider, for example, the argument based on the dynamics of flux tubes.   It is well known that flux tubes break thereby spoiling their pure stringlike dynamics---and the natural connection to a Hagedorn spectrum.   However, it has also long been known that the rate of  flux tube breaking is proportional to the length of the flux tube and inversely proportional to $N_c$ \cite{CasherNeubergerNussinov}.  Thus, in the large $N_c$ limit, flux tubes are arbitrarily long lived and can have simple stringlike dynamics and a Hagedorn spectrum.  The analysis based on gauge-gravity duality ${\cal N}=1 $ super Yang-Mills requires large $N_c$ in its construction and the treatment of  QCD with adjoint fermions in 1+1 dimensions explicitly uses large $N_c$.  Similarly, the analysis in~\cite{Cohen2010, CohenKrejcirik} depends critically on the large $N_c$ limit in numerous ways.

The large $N_c$ limit of QCD is a reasonable caricature of the real world with $N_c=3$ for many phenomena.  However,  it is not always so.  For example, in the large $N_c$ limit, the $\eta'$ meson becomes degenerate with the pion \cite{Witten}, while in the real world it is more  massive than the nucleon.  Another example is the nature of the QCD phase transition at zero chemical potential which is first order at large $N_c$ \cite{LuciniTeper1, LuciniTeper2, Panero} but is only a cross-over at $N_c=3$ and realistic quark masses.  Moreover, the existence of a Hagedorn spectrum depends on the behavior of the spectrum in a limit---the limit of large masses---and one might worry that the large $N_c$ limit and the large mass limits do not commute.  If that is true, the fact that QCD has a Hagedorn spectrum at large $N_c$ need not  tell us anything about the asymptotically high spectrum for finite $N_c$.    So, it is important to see if one can verify the Hagedorn conjecture without reliance on theoretical tools valid only in the large $N_c$ limit.  The large $N_c$ limit was introduced into the problem to simplify the quite complicated theoretical problem of describing hadronic physics.  Thus, it is natural to ask whether one can establish the Hagedorn spectrum---or at least provide solid support for it---from experimental data alone and thus bypass the complicated theoretical issues which prompted the large $N_c$ analysis in the first place.

\section{The Hagedorn spectrum}

 Traditionally, the Hagedorn spectrum was specified in terms of a density of hadron states (perhaps limited to some particular class of hadrons) as a function of mass, $\rho(m)$,  whose definition requires some type of smearing.  The number of hadronic degrees of freedom  with mass between $m$ and $m+ \Delta m$ is given by $\rho(m) \Delta m$ where a hadron of intrinsic angular momentum $J$ is counted as having $2J+1$ degrees of freedom.   The Hagedorn spectrum was one for which  this density asymptotes  at large $m$ to
 \begin{equation}
 \rho_{\rm Hagedorn}(m) = f(m) \exp \left ( \frac{m}{T_H} \right)  \;,	
 \label{Hag1}
 \end{equation}
  where $T_H$ is the parameter known as the  Hagedorn temperature  and $f(m)$ is a subexponential function which asymptotes to a power law at large $m$.   However, as noted by Dienes and Cudell \cite{DienesCudell} and Broniowski and Florkowski \cite{BroniowskiFlorkowski}, a more useful quantity in analyzing the data than $\rho(m)$ is the so-called accumulated spectrum:
  \begin{equation}
  N(m) = \int_0^m {\rm d}m' \rho(m')= \sum_{i={\rm hadrons}} (2 J_i+1)  \theta(m-m_i)  \;,
  \end{equation}
  where $J_i$ is the spin of the $i^{\rm th}$ meson and $\theta$ is the Heavyside step function.  Thus $N(m)$ represents the total number of hadronic degrees of freedom (perhaps restricted to some class of hadron) with mass below $m$.  The Hagedorn conjecture can be stated that at  large $m$, $N(m)$ asymptotes to
  \begin{equation}
  N_{\rm Hagedorn}(m) =g(m) \exp \left ( \frac{m}{T_H} \right )
  \end{equation}
  where $g(m)$ is a subexponential function satisfying $f(m)=g'(m)+g(m)/T_h$.

Before proceeding, it is useful to ask which hadrons ought to be included in an analysis.  As noted above, theoretical arguments suggest that at large $N_c$ both mesons and glueballs ought to have Hagedorn spectra with the same $T_H$.   However, it is by no means clear that these theoretical arguments justify  a Hagedorn spectrum for baryons---even at large $N_c$.  Attempts to fit the baryons by a Hagedorn spectrum yielded a different Hagedorn temperature than  for mesons  \cite{BroniowskiFlorkowski, bronflorgloz, Broniowski} and the  underlying origin of this is not well understood.   Accordingly, we will restrict our attention to hadrons with meson quantum numbers.  Note that if glueballs exist, they have the same quantum numbers as flavor neutral mesons and, except in the theoretical world of large $N_c$, mix with them.  Thus, there is no unambiguous way to separate glueballs from mesons in any analysis.  For this reason when we refer to mesons here we mean hadrons with meson quantum numbers including any possible states with large  glueball content.    For concreteness we will only consider nonstrange mesons.  We do this for practical reasons: one expects qualitatively similar behavior for both strange and nonstrange mesons---indeed precisely the same behavior in the flavor SU(3) limit---however, the strange quark mass will displace strange mesons relative to nonstrange ones and thus may influence attempts to extract a Hagedorn temperature from the data.

However, to ensure that the conclusions we reach are not an artifact of the restriction to non-strange mesons only, in section~(\ref{all}) we  extend the analysis made for non-strange mesons to  all hadrons including baryons.   As we show in that section, the inclusion of all hadrons in the analysis does not alter any of the qualitative conclusions.

At first sight, it  appears to be a simple matter to test the Hagedorn conjecture.  Note that
\begin{equation}
 \frac{m} {\log \left(N_{\rm Hagedorn}(m) \right ) } =\frac{1}{ T_H^{-1} +  \frac{\log \left(g(m)  \right ) }{m} } \rightarrow T_H   \;,
\end{equation}
where the arrow indicates the behavior at large $m$.  However, the Hagedorn conjecture is that $N(m)$ asymptotes to $N_{\rm Hagedorn}(m)$ at large $m$.  Therefore, if one defines $T_H^{\rm eff} (m)$ according to
\begin{equation}
T_H^{\rm eff} (m) \equiv\frac {m}{\log \left(N(m) \right ) }  \; ,
\label{THeffDef}\end{equation}
then the  Hagedorn conjecture is equivalent to the statement that $T_H^{\rm eff}(m) $ asymptotes at large $m$ to a constant  value---which we can identify as the Hagedorn temperature.  Note that if the Hagedorn conjecture is correct, then $T_H^{\rm eff}(m) $ asymptotes to the Hagedorn  temperature at large mass regardless of the form of the subexponential prefactor.   So, in principle all that is needed to verify the conjecture is to find  the empirical $N(m)$ at arbitrarily large $m$ and verify that $T_H^{\rm eff} (m)$ does indeed asymptote.

\begin{figure}
\begin{center}

\includegraphics[width=2.8in]{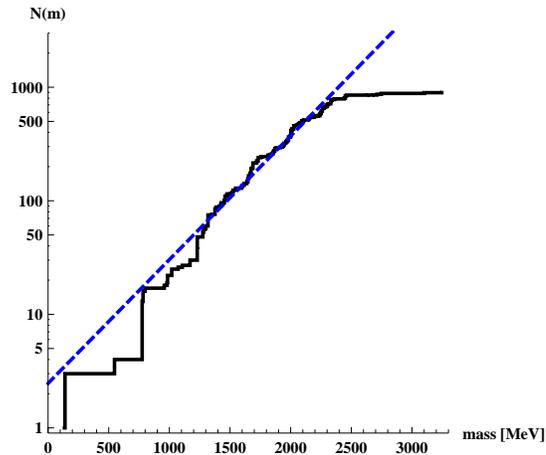}

\caption{Accumulated spectrum of nonstrange mesons plotted logarithmically.  The data are taken from the central values reported by the Particle Data Group \cite{PDG}.  The dashed line represents a simple exponential fit.}
\label{empiricalall}

\end{center}
\end{figure}

Unfortunately this is not practical.   To see why consider the   plot in figure~\ref{empiricalall} of $N(m)$ for nonstrange mesons based on meson masses  from the 2010 report of   Particle Data Group \cite{PDG}.  Up to approximately 2.3 GeV, $N(m)$ appears to be growing essentially exponentially---the data fluctuate  around a simple exponential fit with fluctuations whose relative size becomes small with increasing $m$.  This  behavior is consistent with what one expects from a Hagedorn spectrum.  However, above 2.35~GeV, $N(m)$ bends sharply away from the exponential.  There is a natural explanation for this: As energies of reactions  increase allowing access to higher mass resonances, it becomes increasingly difficult to identify resonances in the typically quite complicated, many-channel partial-wave analysis needed.  Thus, the lack of reported  high mass resonances does not necessarily indicate that they are not present, but only that they cannot be extracted from the scattering data. This means that we cannot use  empirical spectrum at extremely high masses to verify the Hagedorn conjecture with an arbitrarily high level of confidence.

\begin{figure}
\begin{center}

\includegraphics[width=2.8in]{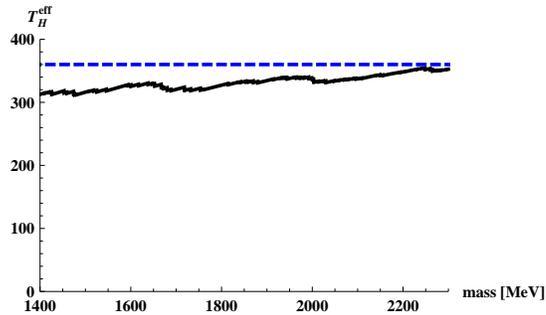}

\caption{$T_H^{\rm eff} (m)$ (as defined in~(\ref{THeffDef})) plotted against $m$.  The dashed line is a constant at 360 MeV and is included to guide the eye.  The meson masses are taken from  the 2010 report of   Particle Data Group\cite{PDG}. }
\label{THeff}

\end{center}
\end{figure}

Nevertheless, the behavior below 2.3 GeV looks to be consistent with the Hagedorn conjecture and provides at least moderately strong evidence for it.   Indeed, the quality of this data has been considered as providing evidence for exponential growth in $N(m)$ \cite{bronflorgloz}.  To make the consistency with Hagedorn conjecture a bit more clear, it is useful to plot $T_H^{\rm eff} (m)$  as defined in~(\ref{THeffDef}) for masses below 2.3 GeV.  This is done in figure~\ref{THeff}.  While it is not completely certain from the data within the range considered that  $T_H^{\rm eff} (m)$ is really approaching an asymptotic value, it seems quite plausible that if we had reliable data  available above 2.3 GeV, $T_H^{\rm eff} (m)$ would asymptote  to a value near 360 MeV.  However, despite this apparent evidence for the Hagedorn conjecture seen in this data, as will be discussed in this paper, the evidence for the Hagedorn conjecture from the meson spectral data is actually quite weak, if there is at all.

Before proceeding, it is important to note that the ``data'' on meson masses included in figure~\ref{THeff} are not really data.  Rather the masses are the results of model-dependent fits.  Model dependence occurs because the hadrons are resonances with non-vanishing widths rather than particles and these resonant structures exist on top of non-resonant backgrounds.  Since we do not have theoretical control of these backgrounds, they must be modeled, and how they are modeled can affect the properties of the resonances including  the mass.  Of course, if the widths are small and the resonances well separated, then  ambiguities due to this model dependence are small and properties of the resonances can be extracted with relatively  small uncertainties.  However, if the resonances are broad and overlapping this is not the case, and the properties of a resonance cannot be pinned down with much certainty from the data---including the critical property of whether or not the resonance exists.  Ultimately, it is precisely this kind of difficulty in extracting resonance properties from scattering data which limits us to relatively low masses in our studies of $N(m)$.

The fact that the masses used in constructing figure~\ref{THeff} are based  on model-dependent fits presents both practical and theoretical difficulties.  At a practical level it raises the question of how to deal with uncertainties in the extracted hadron masses.  The extracted masses have estimated uncertainties \cite{PDG}; however, these estimates include uncertainties due to the model dependence as well as statistical uncertainties and it is not clear how  these should be handled.  Fortunately, if the data are  in the regime where $m$ is large enough so that Hagedorn exponential growth is dominant,  $T_H^{\rm eff} (m)$ should be very insensitive to these uncertainties.  A crude estimate is that
\begin{equation}
\Delta T_H^{\rm eff} (m) \sim \frac{T_H^{3/2} \, \Delta m^{1/2}}{m \,  N^{1/2}} \;,
\label{uncert}
\end{equation}
where $\Delta m$ is the typical uncertainty in the extracted mass.  If we assume that the spectrum is in the Hagedorn regime by $2.3$ GeV and take the typical uncertainty to be around 50 MeV, then~(\ref{uncert}) suggests an uncertainty in $ T_H^{\rm eff}$ of order of 1 MeV.  This is quite small on the scale of $T_H^{\rm eff}$  and can be neglected unless one requires an exceptionally accurate extraction of $T_H$ from the data.  However, as will be shown in detail in this paper, the data are clearly not sufficiently good to demonstrate that the spectrum is in the Hagedorn regime by $2.3$ GeV, and thus it is pointless to attempt a very accurate extraction of $T_H$.  Given this situation we will neglect the uncertainties in the meson masses in all of our analysis.

There is also  a theoretical issue.  Since the Hagedorn spectrum depends sensitively on the properties of mesonic resonances---namely the  existence of the meson and the value of its mass---for high mass resonances, and since these appear to be intrinsically ambiguous with the ambiguity growing with mass, one might worry that the ambiguity due to model dependence renders the question of whether or not a Hagedorn spectrum exists for QCD (for finite  $N_c$)  ill-posed as a matter of principle.   Fortunately, as a mathematical proposition this is not the case.   In principle, {\it  if} one could fully solve QCD, one could compute the correlation function and then analytically continue it for unphysical values of the momenta including complex values.  The analytic structure of these correlators  contains enough information to answer whether a Hagedorn spectrum exists in principle.

 To illustrate this, consider the two-point correlation function for a local composite operator with scalar quantum numbers.  This is a function of one variable, $q^2$.  The analytic structure for $\pi(q^2)$, the correlator, can be quite complicated with multiple cuts associated with $n$-pion thresholds.  However, we know that the ``physical sheet'' only has singularities along the real axis for $q^2>0$.  The strength of this singularity is the spectral function for the correlator up to an overall factor.  Narrow resonances are associated with regions in $q^2$ where the spectral function is large and sharply peaked.  If all we know is information from the physical sheet, we can identify resonances only by some model-dependent fit.   Of course, narrow resonance properties can be determined with little ambiguity.  However, fits to broad resonance will be highly model dependent.

 Suppose, though, one could fully solve the theory including its full analytic structure.  Thus, by hypothesis, we would have access to unphysical Riemann sheets as well as the physical one.  There can be poles on the unphysical sheets.  Moreover, provided that the poles are close to the real axis, their existence greatly constrains the behavior of the spectral function on the physical sheet.  The real part of the pole largely fixes the position in $q^2$ where the spectral function peaks--- {\it i.e.}, the mass of the resonance---and the imaginary part  similarly fixes the width.  Of course, if the poles are not close to the axis, their effect on the pole on the spectral function is far more subtle.  In that case, the pole does {\it not} fix the spectral strength to be a bump centered on the value of the real part with a width determined by an imaginary part.

Given this situation it is natural to {\it define} the resonance mass to be the real part of the associated pole   on a non-physical sheet in the appropriate correlator (or scattering amplitude).  Where there is little ambiguity in the resonance mass as seen from the physical sheet, this prescription reproduces that value (to within the scale of the ambiguity).  Where the ambiguity is large, this prescription resolves the ambiguity in a well-defined manner.  From this perspective, the question of whether QCD has a Hagedorn spectrum is well posed.  Does $N(m)$ grow exponentially where $m$ corresponds to the pole position on an unphysical sheet?

Unfortunately, the definition given above---while making the question well defined in a formal sense---does not aid us in the practical matter of answering the question from scattering data, which by definition, is only directly sensitive to the behavior of the physical sheet.   All that we can do is to look at the mesons whose masses can be determined with relatively small ambiguity and see how these behave and from this attempt to extrapolate to the full function.

With this clarification in mind, let us turn to the  question of whether the data below 2.3 GeV should be taken as strongly supporting the  Hagedorn conjecture.    The first hint that something may be problematic lies in the fact that the apparent value of $T_H$---in the range of 360 MeV---is quite high compared to the value of $\sim 160$ MeV extracted by Hagedorn \cite{Hagedorn2} from hadron-hadron scattering data.   While values of approximately this size \cite{bronflorgloz} and nearly as large \cite{DienesCudell} have been seen in fits to the hadron spectrum, historically most fits have given much smaller values much more consistent with Hagedorn's original value.    For example, a recent extraction by Cleymans and D. Worku  \cite{CleymansWorku} gives a Hagedorn temperature of 174 MeV.   As noted in~\cite{BroniowskiFlorkowski, bronflorgloz}, the differences in the extracted values in the various fits depend sensitively on the functional form of the prefactor to the exponential and the choice of function  typically used historically drops off as $m$ increases and thus biases the fitted  Hagedorn temperature towards smaller values.  Note also that the small value of~\cite{CleymansWorku} is based on a fit including baryons, which---as noted earlier---has little justification.  The large value observed here might be seen as being in tension with the standard picture of Hagedorn.  There is another potential problem with the large apparent value of the Hagedorn temperature seen in this spectral data.   Lattice calculation at relatively large $N_c$  estimates the Hagedorn temperature from a thermodynamic treatment of the metastable hadronic phase above the deconfinement transition (which is first order at large $N_c$) and finds a Hagedorn temperature which is only slightly larger than $T_c$ \cite{LuciniTeper2}.   To the extent that the large $N_c$ world is similar to the world of $N_c=3$, one would expect that the Hagedorn temperature at $N_c=3$ would also be slightly above $T_c$.  However, the Hagedorn temperature extracted from the spectrum here appears to be approximately double the $T_c$.

Given the potential problems associated with the apparently large Hagedorn temperature seen in the meson data, it is useful to ask in detail the extent to which the meson spectral data can be interpreted as strong evidence for Hagedorn conjecture.  Three reasons for caution will be  discussed in this paper.  The first is simply that the rapid increase of $N(m)$ with $m$   seen in the data below 2.3~GeV is well described by simple quark models based on quark-antiquark potentials lacking gluonic degrees of freedom. While Broniowski \cite{Broniowski} previously noted that quark models can describe the rapid rise in $N(m)$, in this paper we observe that even though these models have mesons with arbitrarily high masses, they do not have Hagedorn spectra.    This shows explicitly that the behavior seen in the data does not necessarily imply a Hagedorn spectrum.

A second reason concerns a more detailed analysis of the data.  The region over which the data look to be qualitatively exponential is actually rather small and we show that it is possible to qualitatively reproduce the data with a variety of function forms for the prefactor $g(m)$.  As has been noted elsewhere  \cite{DienesCudell, bronflorgloz, Broniowski}, different  reasonable forms yield quite different values of the Hagedorn temperature indicating that the Hagedorn temperature is not well determined from the data. However, as is stressed in this paper,  if the data were dominated by exponential growth over many e-foldings, the value of $T_H$ would be insensitive to the form of the prefactor.   The fact that it is not implies that the limited data available are not sufficient to conclude that the data are exponential.   Indeed, it is possible to reproduce the data quite well with an infinite Hagedorn temperature, {\it i.e.}, spectrum without exponential growth and thus inconsistent with the Hagedorn conjecture.  Moreover, this can be done in a natural way with a simple power law and does not involve mocking up an exponential by multiple power laws.

In principle,  one could avoid the uncertainty in the Hagedorn temperature if one knew the precise form of the prefactor $g(m)$ from an underlying model. There have been attempts to determine the prefactor from the statistical bootstrap model \cite{Hagedorn1} or a string model \cite{DienesCudell}.
However, the issue under consideration here is not how to extract $T_H$ from the data. Rather, it is whether the data are sufficiently dominated by the exponential growth so that by themselves they provide  strong evidence for the Hagedorn conjecture.

 A third reason is associated with the behavior of the spectrum in various channels associated with fixed meson quantum numbers.   Consistency between a Hagedorn spectrum and the successful Regge phenomenology requires that the channel with the most hadrons grows exponentially with a growth characterized by a Hagedorn temperature equal to that of the whole set.   However, if one looks at a channel-by-channel basis  the data are not consistent with this scenario.  This suggests that the available data are not in a regime dominated by Hagedorn behavior.  Rather, the behavior observed is consistent with  the possibility that much of the rapid increase seen in $N(m)$ is due to the opening of new channels with increasing $m$ rather than the exponential growth within the channels.  While it is easy to see that the effect of opening new channels can by itself only lead to power law growth in $N(m)$, it can also lead to a rapid growth which can be mistaken for exponential growth over a limited range.

\section{Quark models}

For more than four decades quark models---quantum mechanical models based on potentials between quarks---have played an important phenomenological role in describing hadrons.  One of the great ironies in modern physics is that while the quark model was a critical step in the development of the ideas leading to QCD, its connection to  QCD remains quite obscure.  The principle reason for this obscurity is that while QCD has both quarks and gluons as dynamical degrees of freedom, quark models only have quarks.  To the extent that quark models do a phenomenologically adequate job for many purposes, the main effects of the gluon dynamics is somehow encoded in the parameters of the model.

Accordingly, quark models have important intrinsic limitations: effects which probe gluodynamics in detail cannot be described by simple quark models.  Clearly there exist important phenomena in hadronic physics that simple  quark models cannot describe.  An example of this are mesons with ``exotic'' quantum numbers---namely those whose quantum   numbers cannot be obtained in simple quark models with only quark-antiquark degrees of freedom---such as mesons with $I^G(J^{PC})=1^-(1^{-+})$.  In quark model language such states are referred to as ``hybrid mesons'' and are associated with excitations  of ``valence gluons'';  models based on a flux tube  predict such states.   It is known that such states exist and are narrow  in large $N_c$ QCD \cite{Cohen98}.  The Particle Data Group \cite{PDG} includes two exotic  $\pi_1$ mesons with  $I^G(J^{PC})=1^-(1^{-+})$ among the well-established mesons.  There is  good reason to believe that the physics  associated with the Hagedorn spectrum is precisely the kind of gluon-dominated dynamics that simple quark models are incapable of describing.  Consider for example the notion that a Hagedorn spectrum arises from the dynamics of long flux tubes acting like strings.  In that case, the dynamical degrees of freedom---the oscillations of the flux tubes---are associated with the gluon rather than quark degrees of freedom; the flux tubes are gluonic in nature.  Similarly the indirect argument of~\cite{Cohen2010, CohenKrejcirik} is based on a matrix of  correlation functions of an exponentially large number of operators which differ only in their gluonic content.

If it is indeed the case that  simple quark models are not capable of   giving rise to a Hagedorn spectrum, and if such  models {\it are} capable of describing accurately the rapid rise seen in $N(m)$ for meson masses below 2300 MeV, it follows logically that this rapid rise in $N(m)$ need  not imply a Hagedorn spectrum.    As will be shown in this section, there do indeed exist  quark models which describe the rapid rise in $N(m)$ seen in the empirical data and do not have a Hagedorn spectrum.  Thus, we conclude that the data does not, by itself, provide strong evidence for a Hagedorn spectrum.

\begin{figure}
\begin{center}

\includegraphics[width=2.5in]{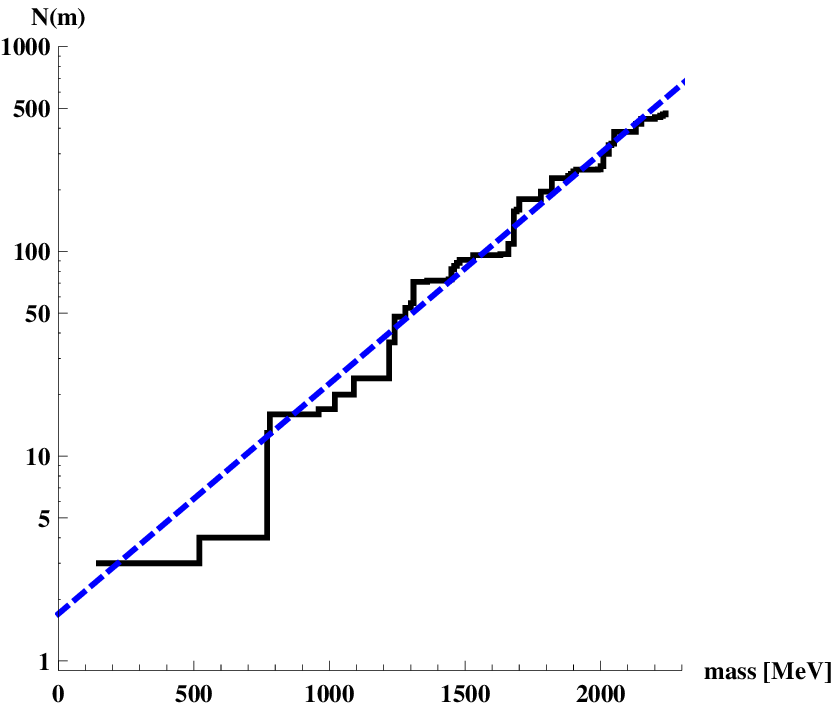} \hskip0.1in
\includegraphics[width=2.5in]{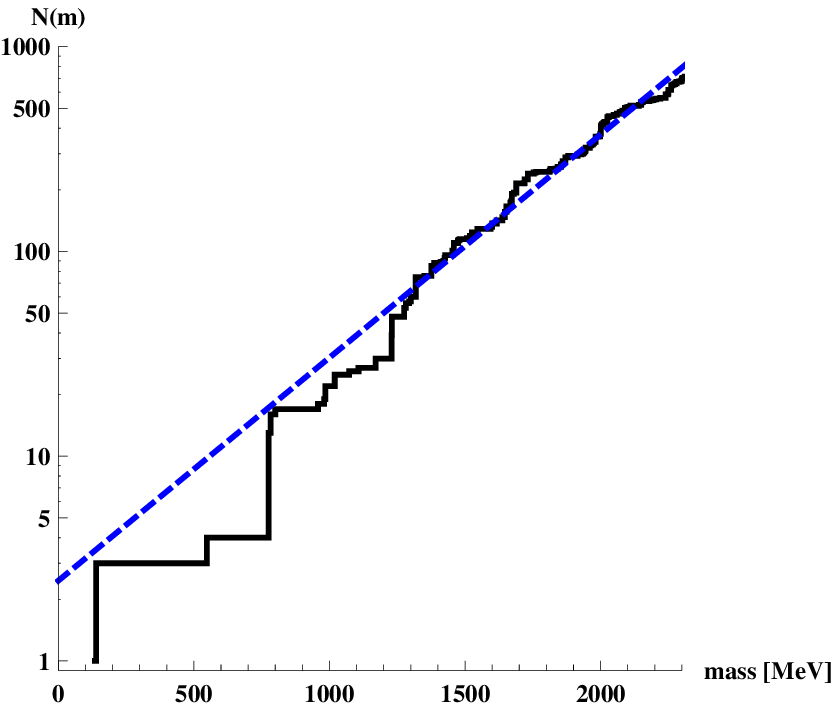}\\

\caption{The accumulated spectrum, $N(m)$ for the GI quark model \cite{GodfreyIsgur} (left figure)  and  for the the empirical spectrum  using data taken from the Particle Data Group \cite{PDG} (right figure).  In both cases the dashed line represents a simple exponential fit.}
\label{GI}

\end{center}
\end{figure}

\begin{figure}
\begin{center}

\includegraphics[width=2.5in]{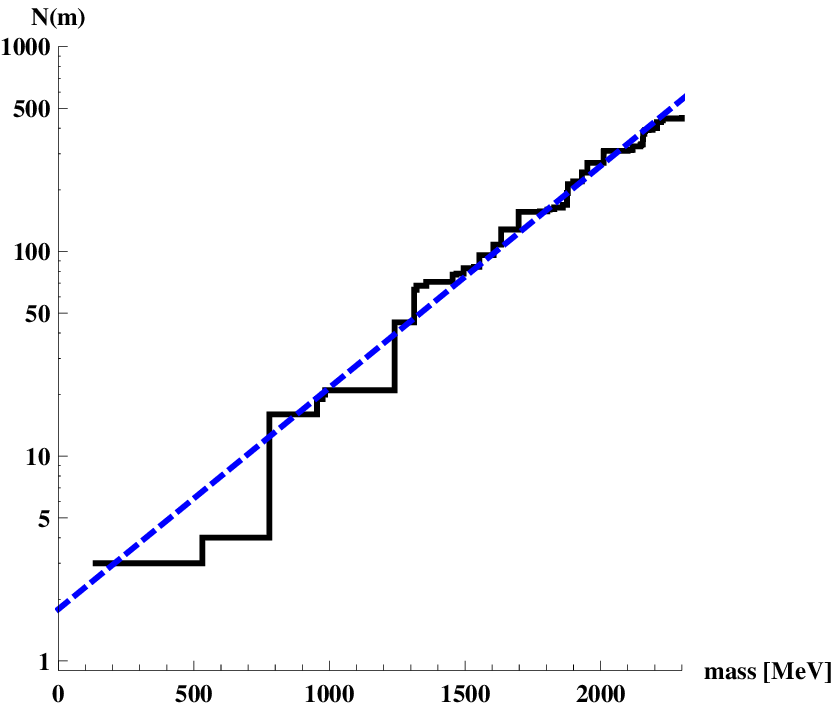} \hskip0.1in
\includegraphics[width=2.5in]{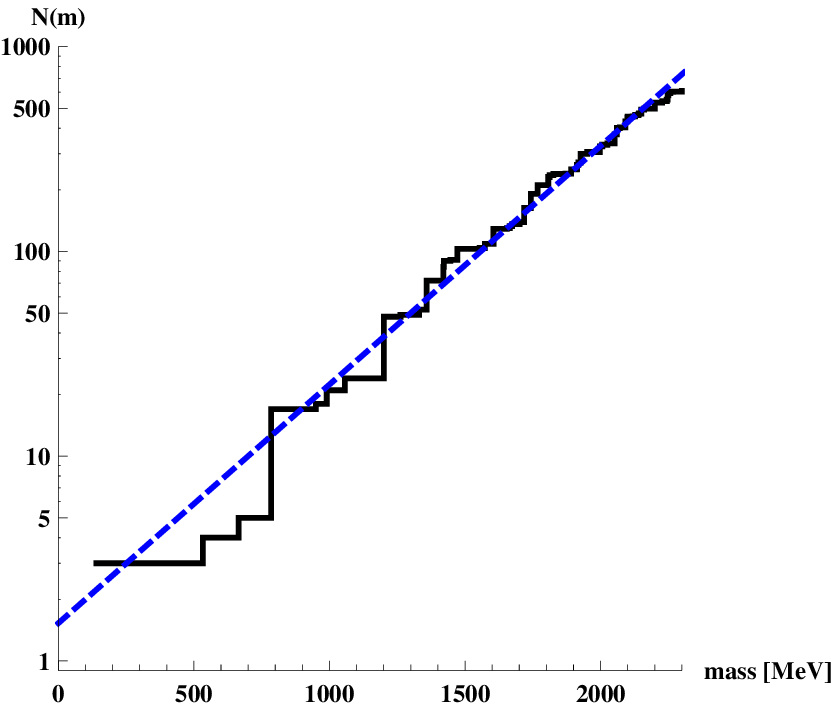}\\
\includegraphics[width=2.5in]{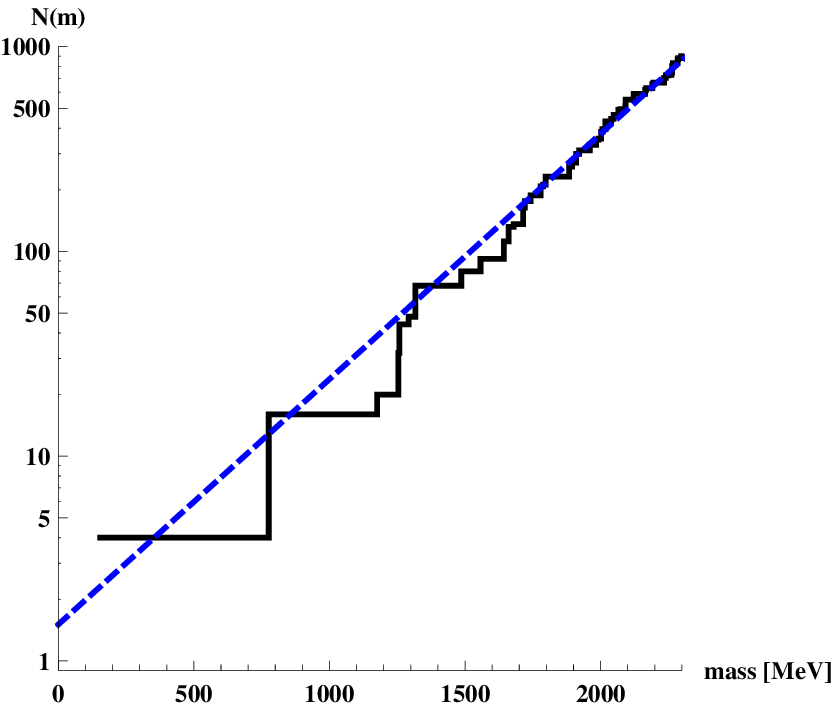}\\

\caption{The accumulated spectrum, $N(m)$ for three quark models.  The top left is for model A from~\cite{Koll}, the top right figure  is for model B of~\cite{Koll} and the bottom figure is for the model in~\cite{EbertFaustovGalkin}.  In all cases the dashed line represents a simple exponential fit.}
\label{others}

\end{center}
\end{figure}

For concreteness, we will illustrate this point by considering in some detail the  classic ``relativized'' quark model of Godfrey and Isgur \cite{GodfreyIsgur}; we shall refer to this as the GI model, which is particularly easy to analyze.      Let us start  looking at the behavior of $N(m)$ for this model.  This is shown in figure~\ref{GI} along with the empirical curve using data from the Particle Data Group.  While it is clear that in detail, the GI model does not precisely reproduce the empirical spectrum, it does have the same qualitative behavior: $N(m)$ rises quite rapidly with $m$ in a fashion which looks to be consistent with a spectrum asymptoting at large $m$ to one with an exponential growth.  There is nothing particularly special about this model from a phenomenological point of view.   As can be seen in figure~\ref{others} the qualitative behavior for $N(m)$ seen in the GI model is also present in other more recent quark models  \cite{Koll, EbertFaustovGalkin}.   Based only on the data  in these plots, it would be quite hard to argue that the empirical data seem to be clearly that of a  Hagedorn spectrum while those of the quark models are not.

However, quark models  can be easily shown not to have a Hagedorn spectrum.   Consider a class of quark models for mesons which has the following features:
\begin{enumerate}
\item The dynamical degrees of freedom in the model are fully characterized by the positions and conjugate momenta  of the quark and antiquark and the spin degrees of freedom of the two.
\item The dynamics is fully specified in some reference frame by a Hamiltonian in terms of these degrees of freedom.
\item The dynamics of the center of mass motion decouples from the internal degrees of freedom.  The internal dynamics is specified by the spins and the position and momenta associated with the relative degrees of freedom and the spin.  The mass of the meson is identified as the energy of the system at zero center of mass momentum.
\item  In the limit where $p$ (the magnitude of the relative momentum) and $\sigma x$ (where $x$ is the magnitude of the relative position, and $\sigma$ is a parameter identified as the string tension) are characteristically  larger than all other quantities in the problem except the energy, the Hamiltonian, is independent of spin and only depends on $p$ and $\sigma x$.   \label{lim}
\end{enumerate}
  Criterion \ref{lim} is particularly important.  It tells us that highly excited states in such models are dominated by the physics of ultrarelativistic quarks  moving in a confining potential.  Thus in this regime the Hamiltonian asymptotes to  a function of the form $h^{\rm asymptotic}( \sigma x, p)$.   The detailed form  of  $h^{\rm asymptotic}( \sigma x,p)$  plays no role in what follows---the critical issue is the parametric dependence.  Clearly the GI model fulfills these criteria.

Since the Hamiltonian is asymptotically fixed by $h^{\rm asymptotic}( \sigma x, p)$ at high energy, it is sufficient to use this form to compute $N(m)$ in the asymptotically high mass regime.  Moreover, in this regime it is legitimate to use the Weyl, semi-classical formula to obtain $N(m)$:
\begin{eqnarray}
N^{\rm asymptotic}(m) &=& 4 N_f^2 \int \frac{ {\rm d}^3 x \, {\rm d}^3 p}{(2 \pi)^3 } \theta \left(  m - h^{\rm asymptotic}(\sigma x, p )  \right )\\
&=& {\rm const}  \frac{4 N_f^2  \,  m^6}{\sigma^3} \;,
 \end{eqnarray}
where $\theta$ represents a unit step function, and the constant depends on the details of  $h^{\rm asymptotic}( \sigma x, p)$.  The important point is that while  $N^{\rm asymptotic}(m)$ grows  quite rapidly with $m$, it grows like a power law---$m^6$---and not an exponential.  Thus quark  models of this type including the GI model do not have Hagedorn spectra.
 Since the GI model does have a rapidly increasing $N(m)$ below 2.3 GeV which is qualitatively  similar to what is seen for the empirical data, it is apparent that this type of qualitative behavior is not by itself strong evidence that the underlying system possesses a Hagedorn spectrum.

\section{The empirical spectrum \label{EmpiricalSpectrum}}

In the previous section we showed that quark models which lack Hagedorn spectra produced rapidly increasing $N(m)$ of a form which was qualitatively similar to that seen in the empirical data below 2.3 GeV.   As suggested above that implies that the qualitative behavior seen in $N(m)$ is not strong evidence for a Hagedorn spectrum.   It is worth noting, however, that quark models have parameters and in one way or another these are chosen to  {\it fit} the hadronic data.   Given this situation, one might wonder whether  the data might really  naturally suggest a Hagedorn spectrum with an exponential growth of $N(m)$ but because the data exists only over a limited domain in $m$  and  has large fluctuations, this exponential behavior is mocked up well by the  non-exponential spectrum of the quark model via the fitting of parameters.  In this section, we test the plausibility of this scenario.

Of course, it {\it is} rather easy to mock up an exponential growth over a limited domain by functions which grow only as  power laws.   For example, consider the function $f(x)=\sum_{n=0}^{10} x^n/n!$.  Over the domain $0 < x < 4$, where $e^x$ increases by more than a factor of 50, $f(x)$ reproduces ${\rm e}^x$ with quite high accuracy; the error at the top to the domain is still under 0.3\%.   On the other hand, a function of the form of $f(x)$ which just happens to reproduce the first 11 terms of the Taylor series of an exponential in almost any circumstance looks to be unnatural---it seems to be a function contrived to mock up the exponential.  The question we wish to address here is whether the fitting of parameters to describe the hadron spectra in effect creates a contrived function.  That is  whether the quality of the data in $N(m)$ is sufficiently exponential in nature that it would take a rather contrived function to mock it up, or, alternatively, does the empirical $N(m)$   naturally accommodate non-exponential fits.  Of course, this question is not a well-posed one mathematically.  Nevertheless it is instructive to look at the empirical spectrum in some detail.

One possible way to proceed is to attempt to fit the empirical $N(m)$ with various  ``reasonable'' functional forms using ``reasonable'' fitting schemes.  One condition on a functional form being reasonable is that it should have few parameters so that it is unlikely  able to mock up the  wrong exponential  behavior by parameter fitting.  If  $N(m)$  really is naturally exponential over a sufficiently large range, one should find that
it is difficult to get a ``reasonable'' fit of the data unless the functional form is $g(m) \exp(m/T_H)$ where $g(m)$  is a  subexponential form.   Moreover, the value of $T_H$ extracted should be fairly similar in all  reasonably good fits to reasonable forms of $g(m)$.  If, on the other hand, one finds that various  good  fits to reasonable functional form yield radically different values of $T_H$ one would be forced to conclude that the available  data do not, by themselves, naturally suggest an exponential behavior characterized by a fixed Hagedorn temperature. The point here is simply that if the growth in $N(m)$ is dominated by the exponential growth over the range of available data, one should be able to pin down the properties of the exponential growth with high accuracy.   If, on the other hand, the growth of $N(m)$ is not dominated by exponential growth it is hard to conclude from the data that the growth is indeed likely to be exponential in character.
 Similarly, if  there are  reasonably good fits to the data using only reasonable  functional forms which grow slower than exponentially, one would again be forced to conclude that the data by itself does not naturally suggest exponential behavior.

\begin{figure}
\begin{center}

\includegraphics[width=2.5in]{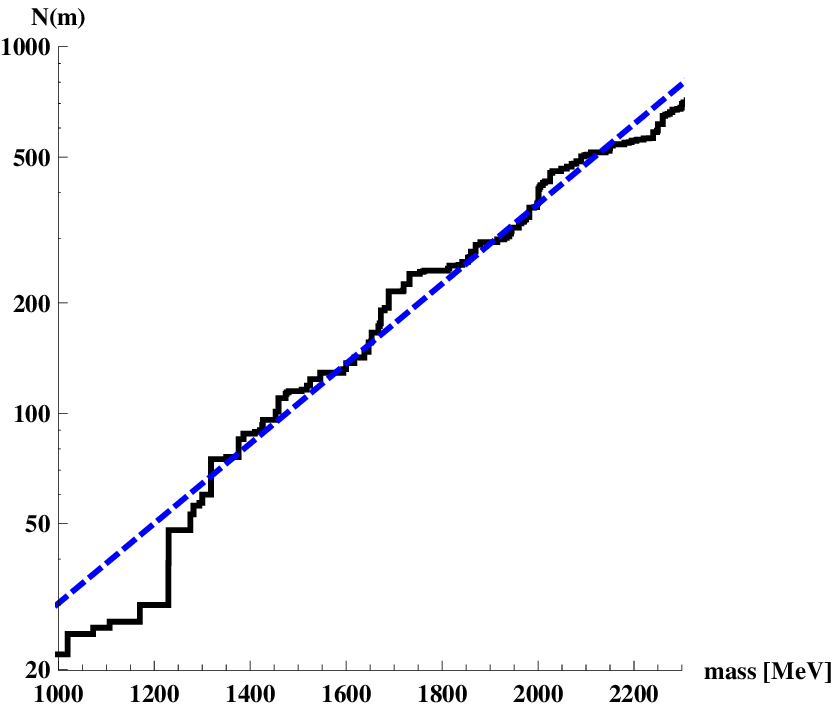} \hskip0.1in
\includegraphics[width=2.5in]{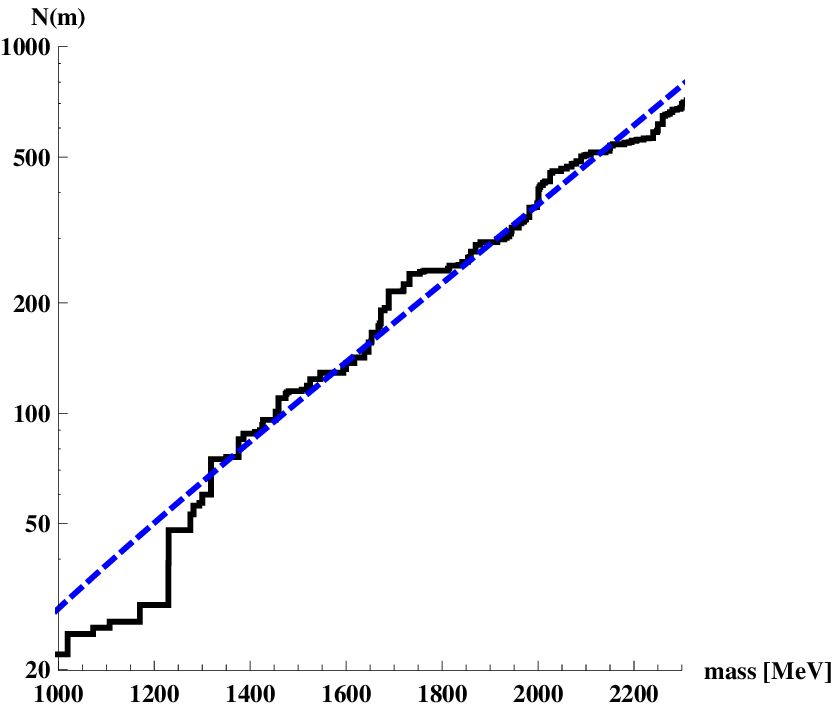} \\ \vskip0.1in
\includegraphics[width=2.5in]{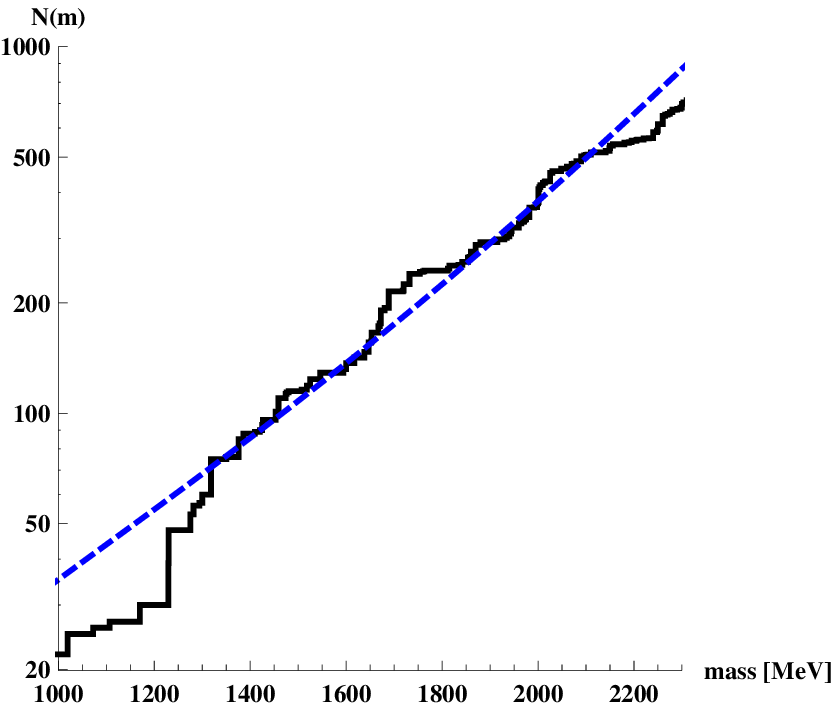} \hskip0.1in
\includegraphics[width=2.5in]{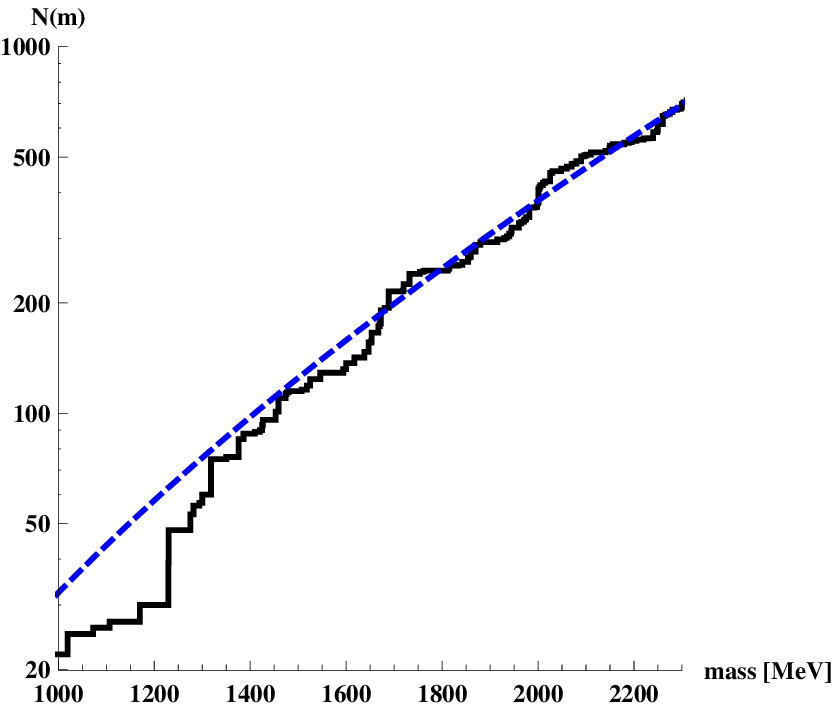} \\

\caption{Four fits to the empirical $N(m)$.  The upper left figure represents fit a) of table \ref{fourfits}.  The upper right figure represents fit b).   The lower left figure represents fit c). The lower right figure represents fit d).}
\label{fits}

\end{center}
\end{figure}

 Before proceeding further it is useful to remark that there is no rigorous way to define a ``reasonable''  functional form in this context.  Similarly there is no obvious way to define ``reasonably good'' fit.  This difficulty of quantifying the quality of the fit has at its origin in the fact that the Hagedorn conjecture is a  statement about the spectrum at arbitrarily large mass and we have no real theoretical control on how rapidly the spectrum should approach its asymptotic form and thus no way to quantify  the size of any systematic errors in the fit.
 Despite this, we can learn something by comparing different fits to the data.  Consider the four fits to the data shown in figure~\ref{fits}.    We have chosen to plot these only for $m$ greater than $1.0$ GeV since our interest is in describing the behavior at large masses.  These correspond to the first four fits listed in table \ref{fourfits}.   All four of them can be said to be ``reasonably good'' given the fluctuating nature of the data and the limited range of the plots.  While fit d)  is probably the qualitatively best fit to the data, fits  a), b) and c)  still appear to describe the data well enough that it would be hard to argue on phenomenological grounds that the physics associated with these  three fits are clearly disfavored.  Note that all four fits use functional forms with only two parameters and thus the prospect of accidentally building a rather contrived functional form through parameter fitting is greatly reduced.

\begin{table}

\caption{Five fits of $N(m)$.}
\label{fourfits}

\begin{indented}

\item[]\begin{tabular}{ c  c  c  c }
\br
                            fit  & functional form & parameter       & parameter           \\\mr
a)     &  {\large\phantom{|}}$A \exp(m/T_H)$    & $A=1.67$        & $T_H= 369$ MeV           \\  & & & \\[-8pt]
b) & $\int_0^m {\rm d}m'  \, \left(\frac{ A }{m'}\right) \, I_2(m'/T_H) $ & $A$= 33.1 &   $T_H=324$ MeV  \\ & & & \\[-9pt]
c) & $\int_0^m {\rm d}m'  \, \frac{ \mu^{3/2} \,\exp(m'/T_H) }{\left ( m'^2+  (500 {\rm MeV})^2 \right )^{5/4}} $ & $\mu$= 1427 MeV &   $T_H=244$ MeV\\
d)     &  $\left (\frac{m}{\mu} \right)^2 \exp(m/T_H)$  & $\mu=326$ MeV         &  $T_H= 876$ MeV                      \\
e)     &   $\left (\frac{m}{\mu}  \right)^p $      & $\mu=470$ MeV        &          $p = 4.11$          \\
\br
\end{tabular}

\end{indented}
\end{table}

 Fit a)  uses  a  pure exponential functional form with a constant prefactor.  The value of the Hagedorn temperature extracted from this fit is 369 MeV.  This is quite close to the value of  $ T_H^{\rm eff} (m)$  from the top end of the data range.  Recall that we had   estimated  a value of $T_H$  in the range of  360 MeV from extrapolating  $ T_H^{\rm eff} (m)$.  At first sight, the  qualitative agreement between $T_H$ in this fit and the value obtained from $ T_H^{\rm eff} (m)$  appears to support the notion that behavior really is Hagedorn-like.  The constant prefactor obtained in the fit is natural---it is a number of order unity which again is consistent with the idea that the underlying spectrum is of the Hagedorn sort.

 Fit b) uses a prefactor derived by Dienes and Cudell \cite{DienesCudell} using a string theory approach. The value of the Hagedorn temperature we obtained using this form (324 MeV) is comparable to the value obtained for a pure  exponential (approximately 10\% smaller). Given only these two fits, one might be tempted to conclude that the data strongly support the Hagedorn conjecture. However, as will be discussed in the following paragraphs,
 the other fits strongly undercut this support.

 Fit c)  uses  the form introduced long ago in~\cite{Hagedorn1} and which has been commonly used  in fits to obtain the Hagedorn temperature. However, we must stress that the functional form used in this fit was motivated  by the statistical bootstrap model developed by Hagedorn in the mid-'60s. This is potentially problematic in that it is not clear the extent to which the statistical bootstrap model is consistent with  QCD, the underlying  theory of strong interactions. Given this, one should be very careful when making statements which depend sensitively  on the detailed form used in this fit.
The value of the Hagedorn temperature extracted from the fit c) is 244 MeV.  Note that the value of the Hagedorn temperature in fit a) is significantly larger than this; indeed it is 50\%  larger. Our value of $T_H$ in this fit is slightly larger  to the value obtained using the same functional form in~\cite{bronflorgloz}; the difference is presumably due to a somewhat different fitting procedure and a slightly different data set.  The value of  the Hagedorn temperature is very different from the fit a); this strongly suggests that the data by itself does not naturally imply an exponential behavior.   The fact that the functional form of the prefactor strongly influences the value of $T_H$ has been observed in past attempts to extract $T_H$ by fitting to the spectrum \cite{DienesCudell, bronflorgloz}.  We stress here that this sensitivity tells us that the data are not in a regime  dominated by exponential growth over a large enough range of masses to clearly distinguish exponential growth from power law growth.

 Fit d) further reinforces this conclusion; it  uses as a  functional form for $N(m)$  a quadratic times an exponential.  The fit to the data is qualitatively quite good---indeed by eye it is clearly superior to fits  a), b) and c).  Moreover, the parameters obtained in the fit seem natural for a hadronic spectra;  they have the dimension of mass  and are in  hundreds of MeV.  The most significant point, however, is that the Hagedorn temperature extracted from this fit is 876 MeV---more than double the Hagedorn temperature extracted from fits  a) and b) or extracted from $ T_H^{\rm eff} (m)$   and more than triple the Hagedorn temperature extracted in fit c).

 Before proceeding, it is useful to understand why the different fits give such different values for $T_H$.  Let us compare fit d) with fit a).  Note that the prefactor in fit d) increases by a factor of $\sim 5.3$  over the range in $m$ from 1000 MeV-2300 MeV.  Also note that over this range the empirical $N(m)$ increases by a factor of $\sim 31.7$.  With this functional form, the exponential growth needed to describe the total growth is a factor of  approximately 6 or $\sim 1.8$  e-folds.   The prefactor thus accounts for almost as large a fraction of the growth as the exponential.   On the other hand, in an exponential fit as in a) one would need the exponential   alone to cause the increase which would amount to $\sim 3.4$ e-folds.  Therefore, if the two fits agreed exactly with the empirical data at 1000 MeV and 2300 MeV, one would have $T_H$ in fit d) nearly double that in  a).  The fact that it is slightly more than double rather than slightly less reflects that the fits are not exactly the same at the top and bottom of the range plotted.  In a similar way, the prefactor in fit and  c) {\it drops} by a substantial amount at the range plotted and thus requires a marked increase in the growth associated with the exponential compared to  a) and hence a  decreased Hagedorn temperature.  A similar, but slightly more complicated analysis could be done for fit b).

\begin{figure}
\begin{center}

\includegraphics[width=2.8in]{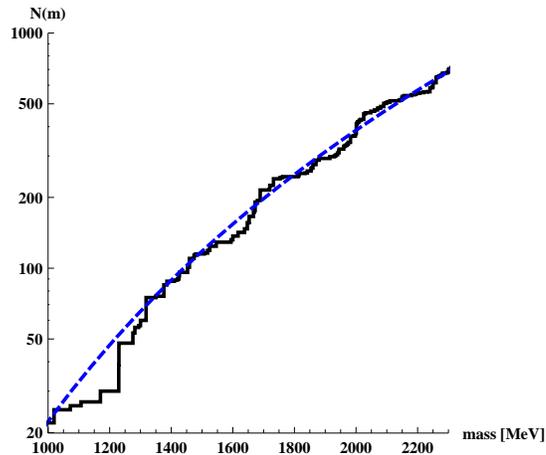}\\

\caption{A fit to the empirical $N(m)$  based on fit e) of table \ref{fourfits}.}
\label{powerfit}

\end{center}
\end{figure}

 That four reasonable fits to the data give very different Hagedorn temperatures strongly suggests  the available  data for $N(m)$ are simply not sufficiently dominated by an exponential growth to deduce that the growth is, in fact, likely to be exponential.   This conclusion is strongly reinforced by fit e) which is a simple power law fit  with no exponential and is shown in figure~\ref{powerfit}.  Note this fit also qualitatively reproduces the data very well.  It also has natural parameters---a mass parameter of several hundred MeV and numerical constant of order one.  The critical issue here is that the fit is to a simple power law.  It appears to fit the data at least as well as the exponential forms in fits a), b), c) and d).    Again we are forced to conclude that the  data by themselves do not naturally suggest an exponential behavior.

This section was focused on the issue of whether the ability of quark models (which lack the exponential growth of a Hagedorn type) to fit the the data on $N(m)$  might reflect little more than parameter fitting of data which in some sense was naturally exponential.  It was shown, however, that the data are not obviously naturally exponential in character.   Various fits with different prefactors give radically different Hagedorn temperatures.  Moreover, a two parameter fit with a single power law explains the data at least as well as a two-parameter fit containing exponentials.  The real lesson appears to be that the data available for $N(m)$ are simply not good enough to decide whether the growth is likely to be exponential at large mass.

\section{A channel-by-channel analysis \label{channel}}

In the previous two sections we demonstrated that the data for mesons with masses less than 2300 MeV, while consistent with $N(m)$ growing exponentially, are also consistent with subexponential growth including the sort one might expect from a quark model.  In this section, we show that if one looks at the data in more detail, the available data appear to be in a regime inconsistent with Hagedorn behavior in which growth of $N(m)$ is dominated by an exponential over many e-folds.

To see why, let us start  by  noting that mesons with various combinations of angular momentum, parity, charge conjugation and isospin are included in $N(m)$.  It is helpful to decompose $N(m)$ in terms of the various  angular momentum-parity-charge conjugation-isospin channels which contribute
\begin{equation}
N(m)= \sum_k N_k(m) \;,
\label{Nk}
\end{equation}
where $k$ represents a particular spin-parity-isospin channel and $N_k(m)$ is the accumulated spectrum for mesons restricted to that channel.   The number of possible channels is infinite as the model supports states of arbitrarily large values of angular momentum.  Of course, at any finite value of $m$ only a finite number of these contributes.   The lowest mass meson with a given $J$ grows with $J$ and thus $N_k(m)=0$ for those channels where $m$ is less than this minimum mass.     New channels open with increasing $m$.

Let us define $n_{\rm ch}(m)$ as the number of channels which contribute ({\it i.e.}, that have nonzero $N_k(m)$ at  mass $m$) and $N^{\rm max}(m)$ to be the largest of the various $N_k(m)$.  By construction
\begin{equation}
 N^{\rm max}(m)  \le N(m) \le n_{\rm ch}(m) N^{\rm max}(m) \;,
\end{equation}
which implies that
\begin{equation}
1- \frac{  \log \left(    n_{\rm ch}(m)    \right ) }{  \log \left(    N(m)    \right ) } < \frac{  \log \left(    N^{\rm max}(m)    \right ) }{  \log \left(    N(m)    \right ) } < 1 \;.
\label{ineq}
\end{equation}

Suppose that the system has a Hagedorn spectrum with an exponentially growing spectrum.  Suppose, further, that we know that $ n_{\rm ch}(m)$  grows no faster than a power law at large $m$ and has ``natural'' sized coefficients.  In such a situation, we know that as $m$ goes to infinity then $\frac{  \log \left(    n_{\rm ch}(m)    \right ) }{  \log \left(    N(m)    \right ) }$ approaches zero and thus $\frac{  \log \left(    N^{\rm max}(m)    \right ) }{  \log \left(    N(m)    \right ) }$ approaches unity.   This means that asymptotically $N^{\rm max}(m)$ must grow exponentially with the exponential growth controlled by $T_H$, the Hagedorn temperature extracted for the full $N(m)$:
\begin{equation}
\lim_{m \rightarrow \infty} \frac{m}{\log \left ( N^{\rm max}(m) \right )} = T_H  \; .
\end{equation}
This is hardly surprising.  Note that in the string picture {\it all} channels grow exponentially with the growth fixed by $T_H$.

\begin{figure}
\begin{center}

\includegraphics[width=2.8in]{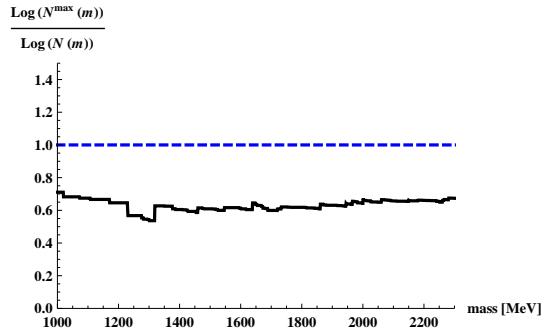}

\caption{The ratio of  $\log \left(    N^{\rm max}(m)    \right )$ to $\log \left(    N(m)    \right ) $ as a function of mass. }
\label{ratio}

\end{center}
\end{figure}

One can use~(\ref{ineq}) to   gauge the extent to which  the data are in ``the Hagedorn regime'' in which the exponential behavior dominates the growth of $N(m)$ over a considerable range of mass.   If $m$ is large enough so that the system is in in this regime and if  the power law growth in $ n_{\rm ch}(m)$  is ``natural'' and contains no anomalously large coefficients then  $\frac{  \log \left(    n_{\rm ch}(m)    \right ) }{  \log \left(    N(m)    \right ) }$ is small and $\frac{  \log \left(    N^{\rm max}(m)    \right ) }{  \log \left(    N(m)    \right ) }$ must be approximately unity.  Thus the extent to which this ratio deviates from unity tell us how far the system is from the Hagedorn regime.  In figure~\ref{ratio} we see that this ratio is well below unity; it is fairly constant with a value of approximately 2/3 for most of the mass range from 1000 MeV to 2300 MeV.  Had both  $N(m)$ and $N^{\rm max} (m)$ been a regime dominated by exponential growth, figure~\ref{ratio} would imply that $N^{\rm max} (m)$ grows much more slowly than $N(m)$, which is inconsistent. Moreover, the ratio in figure~\ref{ratio} shows no evidence of asymptoting towards unity except perhaps at much larger masses. While it is certainly possible that had reliable data been available well above 2300 MeV the ratio would indeed tend toward unity; it seems clear that the regime of $m < 2300$ MeV does not overlap the Hagedorn regime where the growth of $N(m)$ is dominated by the exponential.

\begin{figure}
\begin{center}

\includegraphics[width=2.8in]{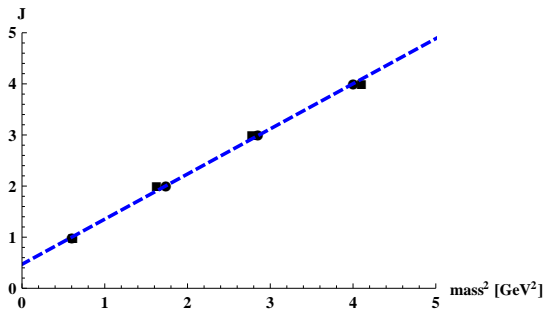}

\caption{The squared mass of the lightest meson with fixed $J$ plotted against $J$.  The squares represent isoscalar mesons and the circles represent isovectors.   Both appear to lie along a Regge trajectory to high accuracy. }
\label{Regge}

\end{center}
\end{figure}

There is a potential caveat to the preceding conclusion.   Note the analysis was based on the assumption that  $n_{\rm ch}(m)$, the number of channels with mesons of mass below $m$, grows as a power law at large $m$.  Before accepting the conclusion we need to verify this assumption.  Note the assumption is correct if the lightest mass meson with fixed angular momentum $J$ falls on a Regge trajectory with the square of the minimum mass  growing linearly with the angular momentum.   This implies that  $n_{\rm ch}(m) \sim m^{2}$ at large mass   since there is a
fixed  number of parity, charge conjugation and isospin possibilities and this number  does not grow with $J$.    Thus we see that {\it if} the lightest mass states of fixed $J$ lie on a Regge trajectory, our assumption is justified.   We know in a string picture, assumed to be valid for high-lying excitations at large $N_c$, these mesons do indeed lie on a Regge trajectory \cite{Strings}.   The critical question is whether the lowest-lying mesons of fixed $J$ lie on a Regge trajectory to good approximation in the regime of relevance here with $N_c=3$ and masses of less than 2.3 GeV.  Of course, as has been known for decades, the answer to this is {\it yes}: the mesons do in fact lie on a Regge trajectory to high accuracy.  This is illustrated in figure~\ref{Regge}.   This confirms that the assumption that  $n_{\rm ch}(m)$ is not growing exponentially rapidly at least in this  domain and by extension indicates that the available data are not in the Hagedorn regime where the growth is dominantly exponential.

\section{An analysis containing all hadrons \label{all}}

As noted in the introduction, there are strong theoretical arguments for the existence of a Hagedorn spectrum for mesons (and glueballs) separately from baryons.  Since these arguments do not directly apply to the baryon spectrum,  it seems that the most sensible way to test the hagedorn conjecture is via a study of mesons only.  This logic is reinforced by the empirical evidence that the rate of growth in the number of hadrons appears to be qualitatively different for mesons and baryons    \cite{BroniowskiFlorkowski} since this  implies that  a combined analysis might give misleading results.  This is particularly true, given the existence of a threshold effect---the baryon spectrum does not start until nearly 1 GeV which may further distort a combined analysis.  Given these circumstances the best test of Hagedorn's conjecture is presumably for the meson spectrum as was  done here.

There is a possible objection to this procedure. One might worry that Hagedorn's conjecture only holds when {\it all} hadrons including baryons  are included; after all Hagedorn's original analysis included all hadrons.   Such a concern is  misplaced.  This can be seen by an argument similar to that of  section~\ref{channel}, but much simpler.  Suppose that we separately tally the number of meson and baryons.  Clearly the accumulated spectrum for hadrons is simply the sum of the accumulated spectrum for mesons and the accumulated spectrum for baryons:
\begin{equation}
N_{\rm hadrons}(m) = N_{\rm mesons}(m) + N_{\rm baryons}(m) \; .
\end{equation}
Now clearly the {\it only} possible ways that $N_{\rm hadrons}(m)$ can  grow exponentially with $m$ is for $ N_{\rm mesons}(m)$ to grow exponentially or for $ N_{\rm baryons}(m)$ to grow exponentially or for   both to grow exponentially.  One can trivially exclude the possibility that the Hagedorn conjecture is true for all hadrons but  holds for neither  the baryon spectrum nor meson spectrum separately.

Despite this, one might nevertheless ask the question of whether the analysis done for mesons would yield qualitatively different results when applied to the spectrum of all hadrons.   One possible advantage to doing such a combined analysis is that it  includes more data.  Since the  principal point of this paper is that the available meson data is not sufficient by itself to give compelling evidence for Hagedorn behavior, that a larger data set might improve things.  Accordingly, in this section, we repeat the analysis done in  sections~\ref{EmpiricalSpectrum} and \ref{channel} with all species of hadrons taken into account.  However, we find that the qualitative behavior is very much the same as with non-strange mesons:  the data is consistent  with the Hagedorn conjecture but the support for it   is  quite weak.  In particular, the data are available over too limited a range to demonstrate convincingly that the data is in a regime dominated by exponential growth.

\begin{figure}
\begin{center}

\includegraphics[width=2.8in]{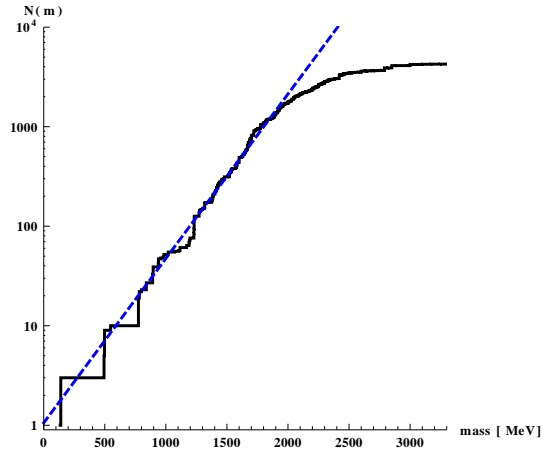}

\caption{Accumulated spectrum of all hadrons plotted logarithmically.  The data are taken from the central values reported by the Particle Data Group \cite{PDG}.  The dashed line represents a simple exponential fit. }
\label{allhadrons}

\end{center}
\end{figure}

In figure~\ref{allhadrons} we have plotted the accumulated spectrum.   Let us note that the significant bend from the straight line in the logarithmic plot (representing exponential growth) occurs already at $\approx$ 2.0 GeV. As noted earlier, such a bend is easily explainable by an increasing difficulty of extraction of high-lying states' properties.    Given this,  our analysis of all hadrons must be limited to an even smaller region of masses  (1.0-2.0 GeV) than the analysis of only non-strange mesons  (1.0-2.3 GeV), where the bend is observed at a somewhat higher mass. Thus, the analysis in this section  will not only be limited to a smaller region of masses than done for meson in sections~\ref{EmpiricalSpectrum} and \ref{channel}, but also is likely to be  distorted by the presence of various types of particles (strange, charmed, baryons, mesons, \dots). The most obvious effect is the baryonic threshold around 1 GeV and their much faster growth rate.

\begin{figure}
\begin{center}

\includegraphics[width=2.2in]{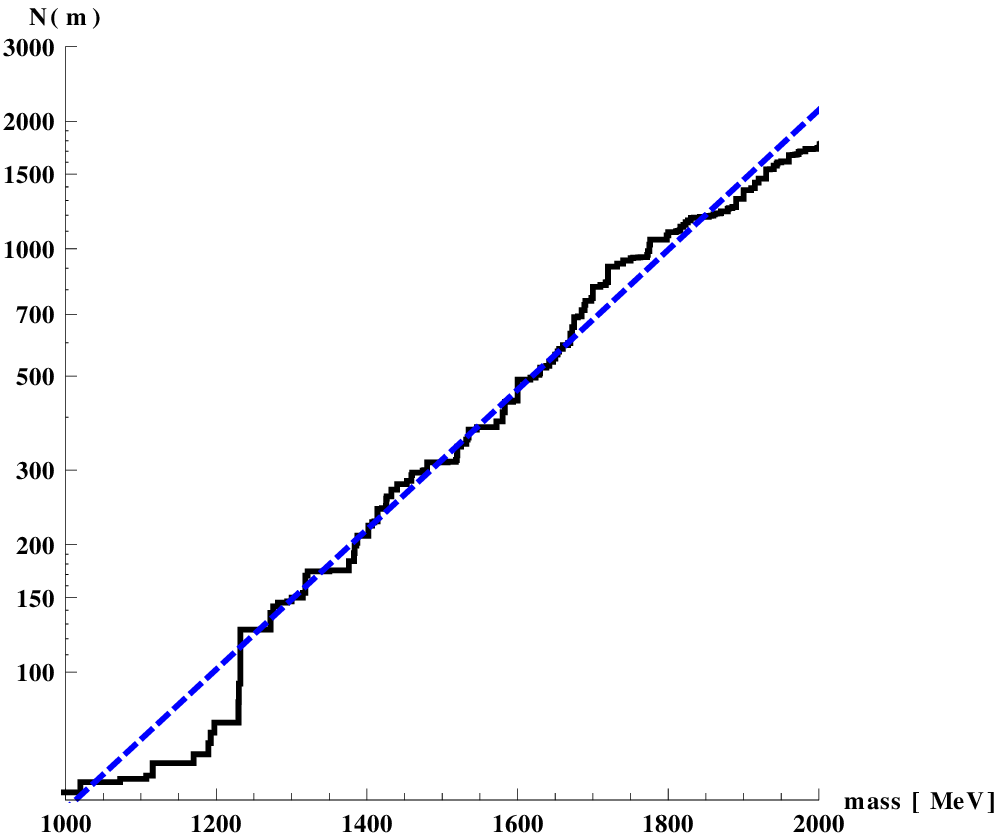} \hskip0.1in
\includegraphics[width=2.2in]{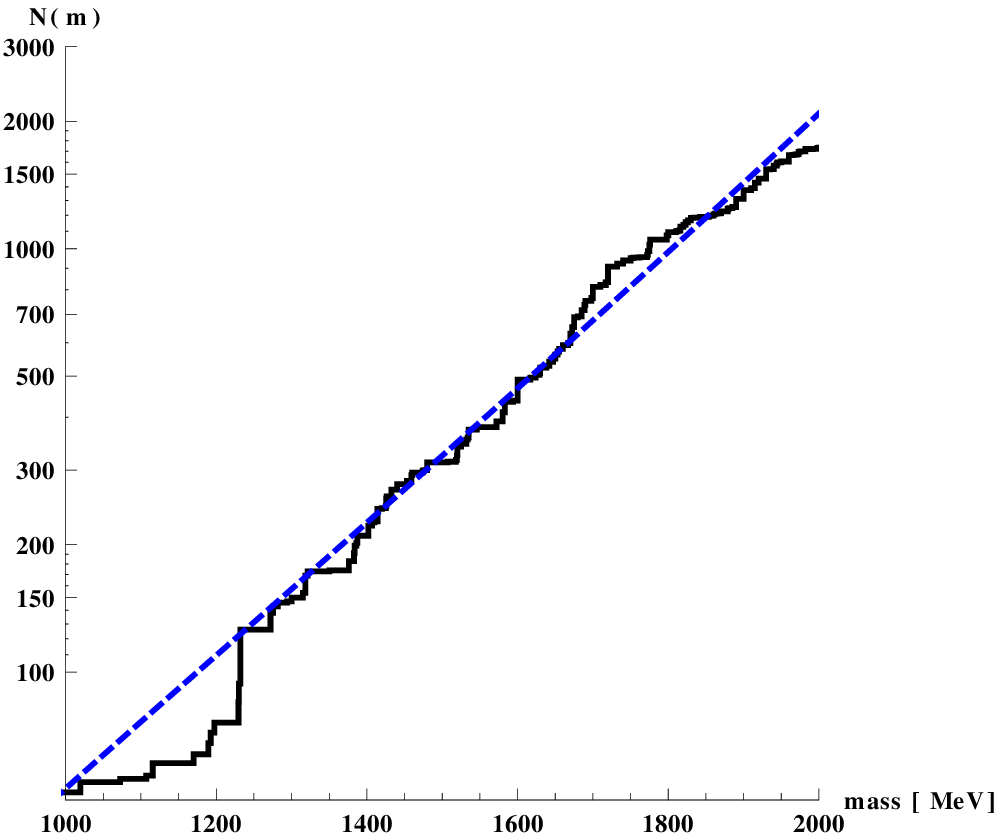} \\ \vskip0.1in
\includegraphics[width=2.2in]{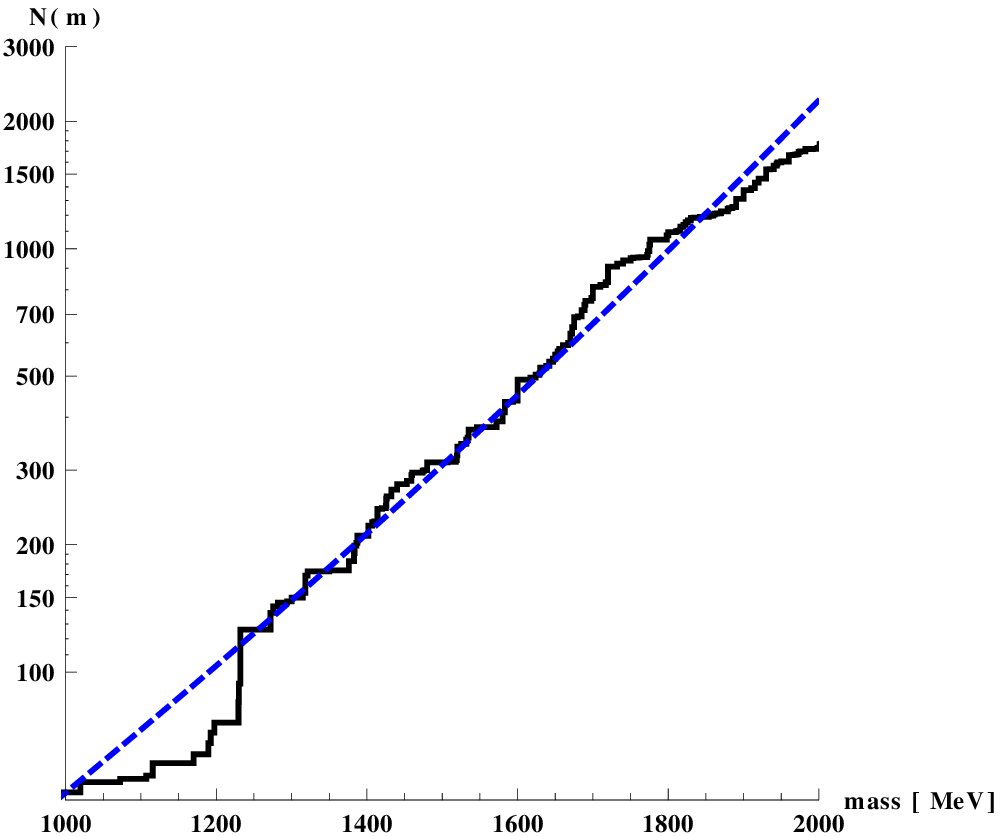}  \hskip0.1in
\includegraphics[width=2.2in]{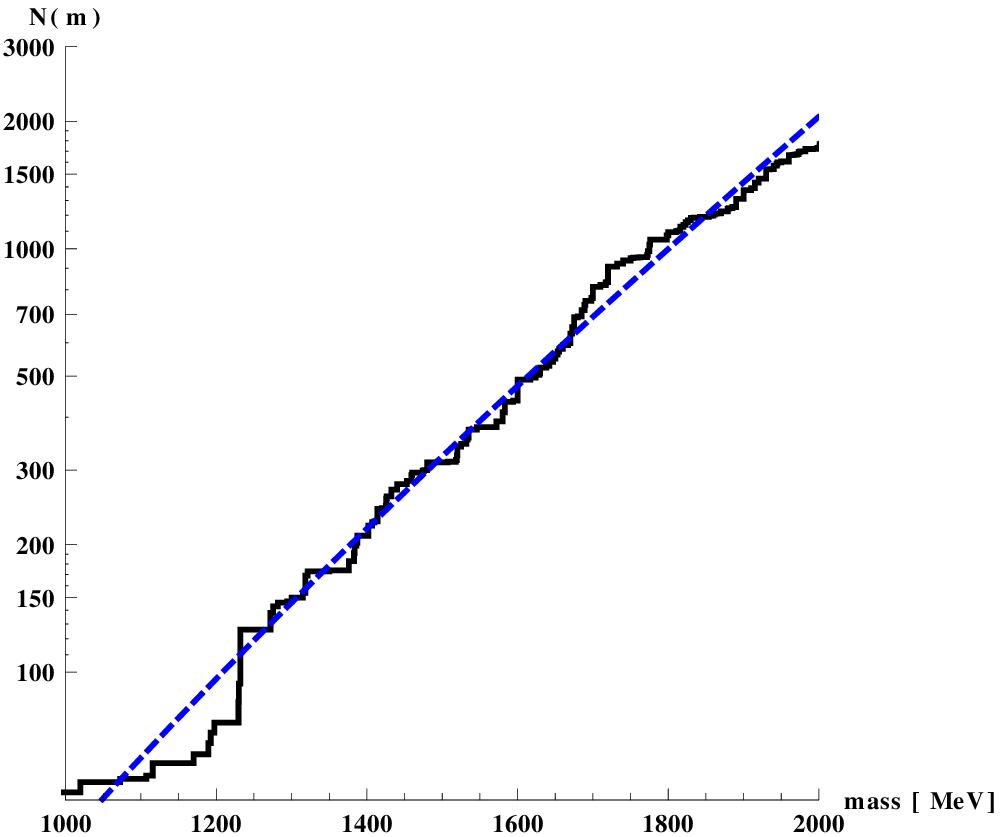}  \\\vskip0.1in
\includegraphics[width=2.2in]{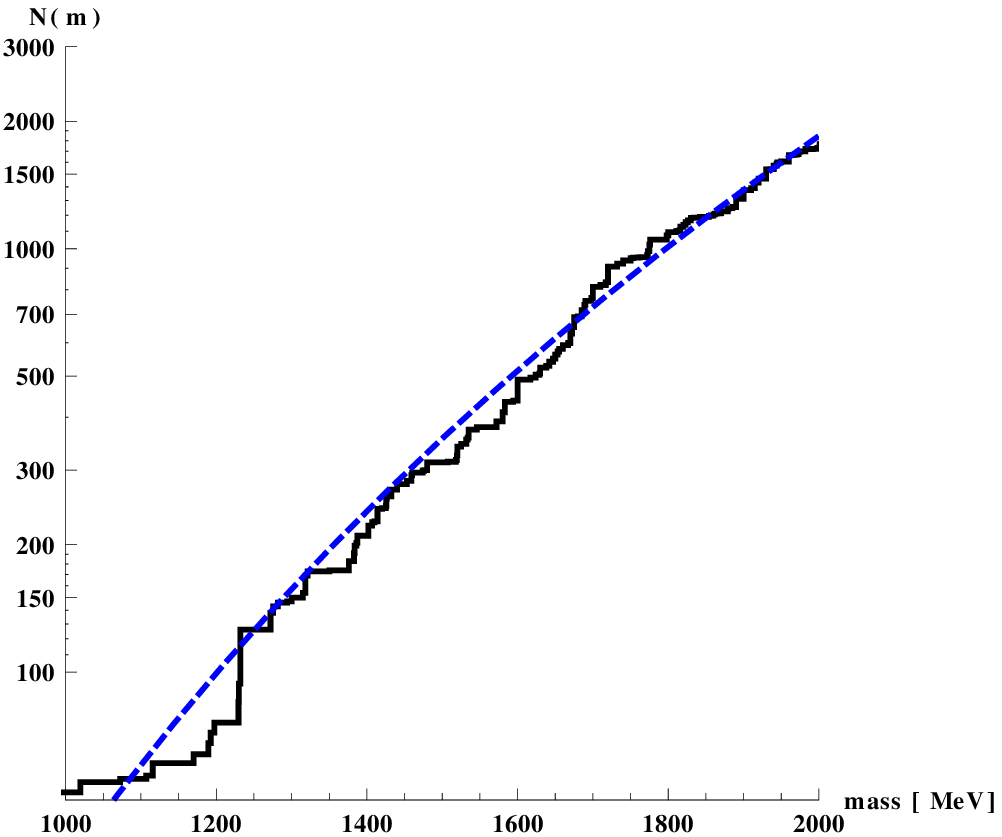} \\

\caption{Five fits to the empirical $N(m)$.  The upper left figure represents fit a) of table \ref{fourfits}.  The upper right figure represents fit b).   The middle left figure represents fit c). The middle left figure represents fit d). The lower right figure represents fit e). }
\label{allhadronfits}

\end{center}
\end{figure}

The fits to the accumulated spectrum of all hadrons were done as they were in section~\ref{EmpiricalSpectrum} and
are summarized in   table \ref{allhadronfourfits} and in figure~\ref{allhadronfits}.
The situation is qualitatively  similar to the case of non-strange mesons only. Once again, one cannot  definitively  favor any of the fits based on direct comparisons with the data; all of them reproduce the data well enough so that discrepancies can easily be accounted for by the limited range of data and their uncertainties. However, the various fits lead to Hagedorn temperatures which differ greatly: ranging from 180 MeV to 390 MeV.  A difference by more than a factor of two {\it in an exponent} is a large difference indeed and suggests that the data is over too small a range to conclude the behavior is exponential.  The fact that a pure power law fit  describes the data as well as any of the exponential fits reinforces this point. Using the same argument as in the section~\ref{EmpiricalSpectrum}, we are forced to conclude that the data  set is not clearly in the region dominated by the exponential growth, i.e., in the Hagedorn regime.

\begin{table}

\caption{Five fits of $N(m)$.}
\label{allhadronfourfits}

\begin{indented}

\item[]\begin{tabular}{ c  c  c  c }
\br
                            fit  & functional form & parameter       & parameter           \\\mr
a)     &  {\large\phantom{|}}$A \exp(m/T_H)$    & $A=1.06$        & $T_H= 262$ MeV           \\
& & &\\[-8pt]
b) & $\int_0^m {\rm d}m'  \, \left(\frac{ A }{m'}\right) \, I_2(m'/T_H) $ & $A$= 18.1 &   $T_H=221$ MeV  \\ & & & \\[-9pt]
c) & $\int_0^m {\rm d}m'  \, \frac{ \mu^{3/2} \,\exp(m'/T_H) }{\left ( m'^2+  (500 {\rm MeV})^2 \right )^{5/4}} $ & $\mu$= 973 MeV &   $T_H=182$ MeV\\
d)     &  $\left (\frac{m}{\mu} \right)^2 \exp(m/T_H)$  & $\mu=562$ MeV         &  $T_H= 393$ MeV                      \\
e)     &   $\left (\frac{m}{\mu}  \right)^p $      & $\mu=537$ MeV        &          $p = 5.72$          \\
\br
\end{tabular}

\end{indented}
\end{table}

\begin{figure}
\begin{center}

\includegraphics[width=2.8in]{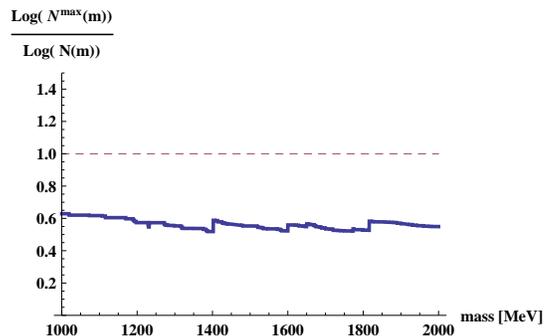}

\caption{The ratio of  $\log \left(    N^{\rm max}(m)    \right )$ to $\log \left(    N(m)    \right ) $ as a function of mass when all hadrons are taken into account. }
\label{allhadronsratioratio}

\end{center}
\end{figure}

The analysis of growth in individual spin-parity channels is also consistent with the statement that the data are outside of the Hagedorn regime where the growth is dominated by the exponential. In section~\ref{channel}, we showed that the ratio of the logarithm of the number of particles in the most populous channel to the logarithm of the number of all particles should reach one in the Hagedorn regime where the growth was dominated by the exponential, provided that the number of open channels does not grow exponentially.  We also noted that if the spectra satisfied Regge trajectories the growth was {\it not} exponential.   As one can see in  figure~\ref{allhadronsratioratio}, the ratio is not unity and remails more or less constant and approximately equal to 0.6.  This behavior is quite similar to what was observed for the meson spectra. In any event the fact that it differs from unity strongly suggests that the data are {\it not} in the Hagedorn regime. In conclusion, the rapid growth in the number of observed hadrons for masses up to 2.0 GeV does not provide strong evidence for a Hagedorn spectrum and is consistent with a growth  caused by the rapid but power law  growth caused by the opening off various spin-parity channels.

\section{Conclusions}

In this paper we considered the evidence for a Hagedorn spectrum both for non-strange mesons and all hadrons. At first sight, the accumulated spectrum of mesons (see figure~\ref{empiricalall}) appears to strongly support the idea that the growth is exponential. Indeed, this kind of evidence appeared sufficiently compelling for the authors of~\cite{bronflorgloz} to indicate a regime of validity of the Hagedorn hypothesis as seen in the data with an upper bound of approximately 2.3 GeV. However, as we demonstrated in this paper, the spectral data on mesons do not provide the strong evidence for the Hagedorn conjecture. We showed that the behavior is similar if one does  an analysis  of the spectrum of all hadrons too.

We showed mathematically that quark models such as the model of Godfrey and Isgur cannot give a rise to Hagedorn spectrum (and moreover on physical grounds should not have one). And yet, qualitatively, the overall spectra generated by various quark models look as exponential as that seen in the empirical data. We also performed fits to the spectra with various functional forms with mostly two free parameters. Had the spectrum actually been in the Hagedorn regime in which the growth in the number of mesons with $m$ was dominated by exponential growth over many e-foldings, the Hagedorn temperature extracted from different fits should be roughly the same for any reasonable form of the prefactor to the exponential. However, we obtained reasonable fits of the data with Hagedorn temperatures which ranged from 244 MeV to infinity with various reasonable choices of prefactor. The fact that a good fit to the data was seen with $T_H= \infty$,  {\it i.e.}, a pure power law, is a particularly compelling demonstration that the data by itself do not imply a Hagedorn spectrum. The third argument against the experimental evidence of Hagedorn spectrum was based on the channel-by-channel analysis of meson spectrum. In the Hagedorn regime, the spectrum of mesons in the channel with the maximum number of mesons will grow exponentially at the same rate as the full spectrum. Empirically this is not the case.

Thus there is strong evidence that over the available range of $m$, the data are not in the Hagedorn regime. Had the data been in the Hagedorn regime it would have been possible to get strong evidence for exponential growth since the exponential growth dominates. Indeed, in this regime there is no natural way to describe the data without an exponential with approximately the right Hagedorn temperature; the only way to avoid describing the data with an exponential with nearly the correct Hagedorn temperature would be by constructing a very contrived function which mocks up an exponential over the available range of the data. However, away from the Hagedorn regime there is no clean way to see whether the growth in $N(m)$ is really exponentially growing. Since we have shown that the available data are not in this regime, we conclude that the data on the meson spectrum are not of high enough quality to give a meaningful indication of whether or not the Hagedorn conjecture is correct.

We wish to stress that our conclusions are based on rather qualitative behavior and not on a precise quantitative analysis. This approach is necessary due to the quality of the experimental data and its limited range in mass. Given the fact the Hagedorn hypothesis only requires that the spectrum should reach its Hagedorn regime in the asymptotically high mass region, we are, most of all, interested in the high-lying mass states. Yet, this is exactly the region where the quality of the experimental data is the poorest. The principal reason is that the analysis of the actual experimental data becomes increasingly difficult to analyze with increasing energy due to the large number of open channels. Thus, the extraction of resonance properties from the partial wave analysis becomes more model dependent and even the crucial question of whether or not a given resonance exists is not always clear. All these facts forced us to restrict our analysis to mesons with masses less than 2.3 GeV and to rely on criteria based on reasonableness.

For the reasons articulated above, we were careful in section \ref{EmpiricalSpectrum} not to distinguish between various reasonable fits to the data. However, suppose for the moment  we disregard this and take seriously the differences in quality of the various fits. It seems pretty clear that fits  d) and e) from table~\ref{fourfits}, corresponding to the last curve in figure~\ref{fits} and the one in figure~\ref{powerfit},
are qualitatively noticeably better than fits  a), b) and c). The reason for this is pretty clear: if we look more precisely at the logarithmic plot of the accumulated meson spectrum (for example figure~\ref{powerfit}) we see that the curve has a tendency to bend somewhat downward at increasing $m$. Prefactors with positive power laws (such as in fits  d) and e)) automatically have this behavior and hence one expects fits where the prefactors have positive power laws to describe the data somewhat better than a constant prefactor (as in fit  a)) or one with a negative power (as in fits  c)) which curves upward.  Fit b)  uses a more complicated structure---a modified Bessel function instead of a pure exponential---which affects the shape of the curve. Indeed, there is an inflection point, where the curve switches from turning downward to turning upward and seems almost straight. Consequently, the Hagedorn temperature for this fit is qualitatively similar to the   pure exponential.   If  we could rely on the quality of the fits to distinguish between the likely validity of the various functional forms, we would be forced to conclude that the data suggests very large Hagedorn temperatures---fit  d) has $T_H=876$ MeV while fit  e) has $T_H=\infty$.

However, we should not rely on the quality of the fits to distinguish between the likely validity of the various functional forms. Apart from the general concerns articulated above, one can easily imagine that that the qualitative behavior favoring this fits---the fact that a logarithmic plot of $N(m)$ curves downward could be an artifact. As discussed in the introduction, as $m$ increases it becomes increasingly difficult to extract resonances from the scattering data. Thus, it is plausible that with increasing $m$ an increasing fraction of resonances is missed in the analysis of the scattering data. This would have the effect of causing an artificial bend downward in the logarithmic plot of $N(m)$ and would artificially improve the quality of the fits with high---or infinite---Hagedorn temperatures. Given these concerns, we are quite reluctant to argue that the spectral data disfavor exponential growth unless there is a very high Hagedorn temperature. At the same time, the available data clearly do {\it not} favor exponential growth with a low Hagedorn temperature. Indeed, our overall conclusion is simply that the quality of the spectral data and its limited range appears to be inadequate for drawing any conclusions about the nature of the asymptotic behavior of $N(m)$.

In the introduction, stress was placed on the importance of using empirical data on the spectrum to establish the Hagedorn hypothesis for QCD in the physical world with $N_c=3$. There is strong theoretical evidence for a Hagedorn spectrum in large $N_c$ QCD. However, it is by no means clear how these large $N_c$ arguments can be extended to finite $N_c$. Thus, the present situation is that the question of whether QCD with $N_c=3$ has a Hagedorn spectrum is open. There are neither compelling theoretical nor experimental reasons to believe that it does.

\ack

This work was supported by the U.S.~Department of Energy
through grant DE-FG02-93ER-40762.


\section*{References}

\begin{thebibliography}{99}

\bibitem{Hagedorn1}   Hagedorn R 1965 {\it Nuovo Cimento Suppl.}  {\bf 3}  147

\bibitem{Hagedorn2}  Hagedorn R 1968 {\it Nuovo Cimento} {\bf 56A} 1027

\bibitem{HuangWeinberg} Huang K and  Weinberg S 1970 {\it Phys. Rev. Lett.} {\bf 25} 895


\bibitem{CabiboParisi} Cabibbo N and  Parisi G 1975 {\it Phys. Lett. B} {\bf 59} 67

\bibitem{Cohen2006} Cohen TD, {\it Phys. Lett. B} {\bf 637} 81


\bibitem{Strings}  Polchinski J 1998 {\it String Theory} (Cambridge: Cambridge University Press)

\bibitem{BraunMunzinger1} Braun-Munzinger P and Stachel J 1996 {\it  Nucl. Phys. A}  {\bf 606}  320

\bibitem{BraunMunzinger2} Braun-Munzinger P, Redlich K and Stachel J 2004 {\it Quark-Gluon Plasma 3} (World Scientific Publishing) p~491


\bibitem{BraunMunzinger3} Braun-Munzinger P and Stachel J 2011 ({\it Preprint} nucl-th/1101.3167)

\bibitem{IsgurPaton}	Isgur N and Paton J 1985 {\it Phys. Rev. D} {\bf 31} 2910

\bibitem{KoganZhitnitsky}  Kogan II and  Zhitnitsky AR 1996 {\it Nucl. Phys. B} {\bf 465} 99

\bibitem{PandoZayas}  Pando Zayas LA and Vaman D 2003 {\it Phys. Rev. Lett.} {\bf 91} 111602-1

\bibitem{Cohen2010} Cohen TD 2010 {\it JHEP} {\bf 06} 098

\bibitem{CohenKrejcirik} Cohen TD and  Krej\v{c}i\v{r}\'{i}k V 2011 {\it JHEP} {\bf 08} 138

\bibitem{CasherNeubergerNussinov} Casher A,  Neuberger H and  Nussinov S 1979 {\it Phys. Rev. D} {\bf 20} 179-188

\bibitem{Witten} Witten E 1979 {\it  Nucl. Phys. B} {\bf 160}  57

\bibitem{LuciniTeper1} Lucini B, Teper M and  Wenger U 2004 {\it JHEP} {\bf 01} 061

\bibitem{LuciniTeper2} Lucini B, Teper M and Wenger U 2005 {\it JHEP} {\bf 02} 033

\bibitem{Panero} Panero M 2009 {\it  Phys. Rev. Lett.} {\bf 103} 232001

\bibitem{DienesCudell} Dienes KR and Cudell JR 1994 {\it Phys. Rev. Lett.} {\bf 72}  187

\bibitem{BroniowskiFlorkowski} Broniowski W and Florkowski W 2000 {\it  Phys. Lett. B} {\bf 490}  223

\bibitem{bronflorgloz}  Broniowski W,  Florkowski W and Glozman LY 2004 {\it Phys. Rev. D} {\bf 70} 117503-1

\bibitem{Broniowski}  Broniowski W 2000 ({\it Preprint} hep-ph/0008112)


\bibitem{PDG} Nakamura K and Particle Data Group 2010 {\it J. Phys. G } {\bf  37} 075021




\bibitem{CleymansWorku} Cleymans J and  Worku D 2011 {\it Mod. Phys. Lett. A} {\bf 26}  1197

\bibitem{Cohen98} Cohen TD 1998 {\it  Phys. Lett. B} {\bf 427} 348

\bibitem{GodfreyIsgur} Godfrey S and  Isgur N 1985 {\it Phys. Rev. D} {\bf 32} 189

\bibitem{Koll} Koll M, Ricken D,  Merten D,  Metsch BC and  Petry HR 2000 {\it Eur. Phys. J. A} {\bf 9}  73

\bibitem{EbertFaustovGalkin} Ebert D,  Faustov RN and  Galkin VO 2009 {\it Phys. Rev. D} {\bf 79} 114029-1



\end{thebibliography}
\end{document}